\documentclass[sigconf]{acmart}
\AtBeginDocument{%
  }

\copyrightyear{2026}
\acmYear{2026}
\setcopyright{cc}
\setcctype{by}
\acmConference[CHI '26]{Proceedings of the 2026 CHI Conference on Human Factors in Computing Systems}{April 13--17, 2026}{Barcelona, Spain}
\acmBooktitle{Proceedings of the 2026 CHI Conference on Human Factors in Computing Systems (CHI '26), April 13--17, 2026, Barcelona, Spain}
\acmDOI{10.1145/3772318.3791269}
\acmISBN{979-8-4007-2278-3/2026/04}

\usepackage{enumitem}
\usepackage{hyperref}

\begin{document}
\title[Routine Computing]{Routine Computing: A Systematic Review of Sensing Daily Life Dimensions Towards Human-Centered Goals}

\author{Borislav Pavlov}
\authornote{Both authors contributed equally to this research.}
\email{borislavgpavlov@gmail.com}
\orcid{0009-0008-4768-408X}
\affiliation{%
  \department{Key Laboratory of Pervasive Computing, Ministry of Education}
  \department{Department of Computer Science and Technology}
  \department{Beijing National Research Center for Information Science and Technology (BNRist)}
  \institution{Tsinghua University}
  \city{Beijing}
  \country{China}
}

\author{Jiajin Li}
\authornotemark[1]
\authornote{Visiting student at the time of this work.}
\email{lijiajin0516@gmail.com}
\orcid{0009-0001-5723-5383}
\affiliation{%
  \department{Department of Computer Science and Technology}
  \institution{Tsinghua University}
  \city{Beijing}
  \country{China}
}

\author{Jun Fang}
\email{fangy23@mails.tsinghua.edu.cn}
\orcid{0009-0001-2614-8674}
\affiliation{%
  \department{Department of Computer Science and Technology}
  \institution{Tsinghua University}
  \city{Beijing}
  \country{China}
}

\author{Yuntao Wang}
\authornote{Corresponding author.}
\email{yuntaowang@mails.tsinghua.edu.cn}
\orcid{0000-0002-4249-8893}
\affiliation{%
  \department{National Key Laboratory of Human Factors Engineering}
  \department{Department of Computer Science and Technology}
  \institution{Tsinghua University}
  \city{Beijing}
  \country{China}
}
\affiliation{%
  \institution{Ant Group}
  \city{Hangzhou}
  \country{China}
  }

\author{Yuanchun Shi}
\email{shiyc@tsinghua.edu.cn}
\orcid{0000-0003-2273-6927}
\affiliation{%
  \department{Key Laboratory of Pervasive Computing,Ministry of Education}
  \department{Department of Computer Science and Technology}
  \department{BNRist}
  \institution{Tsinghua University}
  \city{Beijing}
  \country{China}
}


\renewcommand{\shortauthors}{Pavlov et al.}

\begin{abstract}
Human routines structure daily life, yet remain challenging for computational systems to understand. This paper presents the first systematic review of routine computing, a previously implicit but increasingly recognized field that focuses on computationally sensing and modeling human behaviors. It synthesizes 203 studies published up to August 2025. The paper presents a new taxonomy of the literature, focusing on temporal structures, behavioral interactions, cognitive aspects, and how variability and deviations are addressed. The common goals of routine computing extend across four major application domains, including accessibility care, the promotion of healthy habits, adaptive and context-aware support, and large-scale population insights. Persistent challenges that limit the design of truly human-centered systems are identified, including the gap between low-level activity recognition and high-level intent, the tension between personalization and generalization, unresolved privacy concerns, and data-related limitations. By consolidating these findings, this paper provides a foundational framework for HCI researchers, outlining principles for designing ethical, adaptive, and human-centered routine-aware systems.
\end{abstract}

\keywords{Do, Not, Use, This, Code, Put, the, Correct, Terms, for,
  Your, Paper}


\maketitle

\section{Introduction}
\begin{figure*}[t]
    \centering
    \includegraphics[width=1.0\linewidth, keepaspectratio]{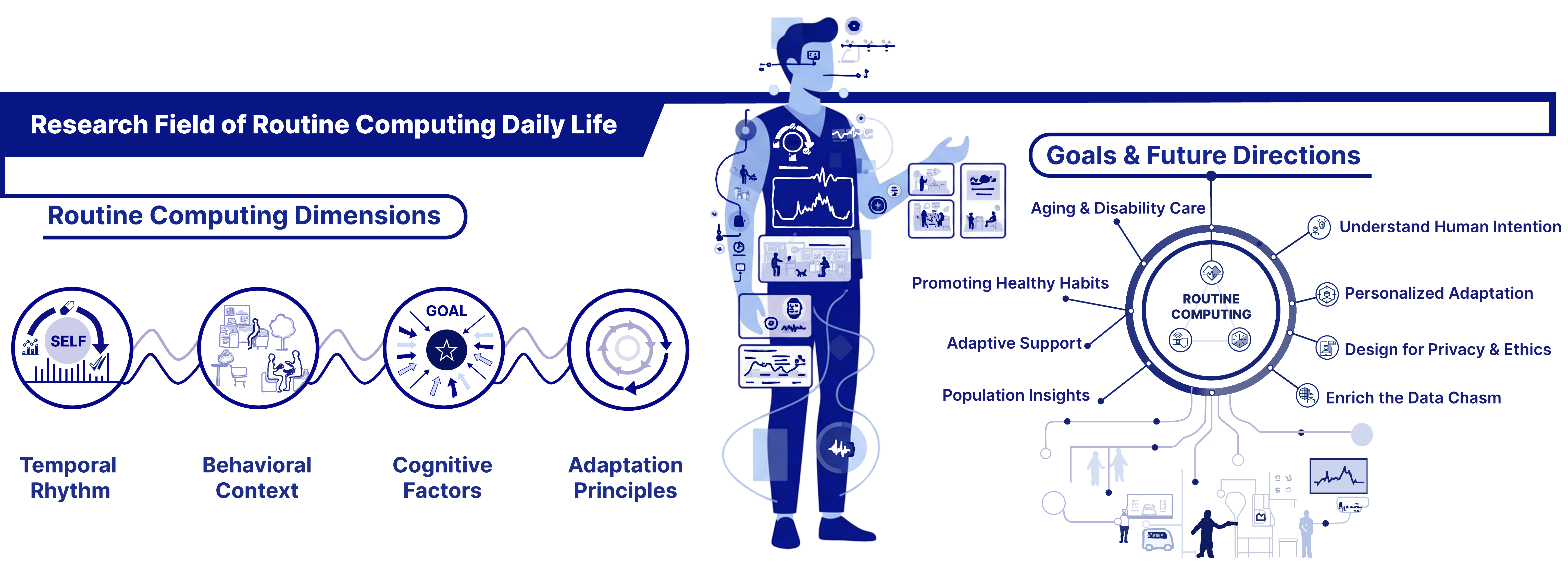}
    \caption{Visual abstract of routine computing survey and taxonomy. This framework summarizes the conceptual landscape of routine computing across three major layers. Dimensions outline four core perspectives for characterizing everyday routines: \textit{Temporal Rhythm}, \textit{Behavior Context}, \textit{Cognitive Factors}, and \textit{Adaptation Principles}. Goals describe four primary application targets, including \textit{Aging and Disability Care}, \textit{Promoting Healthy Habits}, \textit{Adaptive Support}, and \textit{Population Insights}. Future Directions highlight emerging opportunities for advancing the field through \textit{Understanding Human Intention}, \textit{Personalized Adaptation}, \textit{Design for Privacy and Ethics}, and \textit{Enriching the Data Chasm}.}
    \Description{Diagram showing the dimensions of routine computing: sensing granularity, behavioral and social contexts, cognitive and psychological factors, and application domains.}
  \label{fig:teaser}
    \label{fig:cover}
\end{figure*}
Human life is structured by routines—recurring sequences of activities such as commuting, eating, working, and resting—that give rhythm, meaning, and stability to everyday experience. Recent advancements in ubiquitous computing have sought to integrate more seamlessly into these daily patterns by modeling and responding to human behavior. Yet, as Suchman’s \textit{Situated Action} \cite{suchman1987} and Dourish’s \textit{Embodied Interaction} \cite{dourish2001} remind us, human behavior is deeply contextual and interpretive, making the computational modeling of routines particularly demanding. 

The study of human routines in computing draws on several long-standing research traditions. Human Activity Recognition (HAR) research, beginning with foundational work by Katz \cite{katz1963} and Lawton and Brody \cite{lawton19699}, established formal taxonomies of self-care and instrumental tasks that later informed monitoring practices in gerontology, disability care, and home-based health assessment. Context-aware computing \cite{dey1999} extended this perspective by emphasizing that meaningful interaction requires sensing and interpreting the situational factors surrounding an activity—location, timing, social presence, and environmental state. Across these traditions, prior surveys have offered important overviews of related areas \cite{review1, review2}, however, they were not designed to address the broader patterns that structure daily life. They typically focus on momentary or single-category activities, rather than the complex, evolving, and interdependent actions that unfold across hours, days, and weeks. To address this gap, we introduce the concept of \textbf{routine computing}, defined as the computational sensing, modeling, understanding, and application of recurring human behaviors in daily life. Routine computing extends beyond activity recognition by foregrounding the temporal, contextual, and psychological dimensions that give everyday practices their meaning and significance. \\
Our work is guided by the following research questions:
\begin{enumerate}[nosep,label=\textbf{RQ\arabic*:}]
    \item What conceptual dimensions are emphasized by existing studies of daily routines in HCI and ubiquitous computing research?
    \item What goals drive the development of routine-aware technologies?
    \item How do the identified dimensions and goals of routine computing shape the user-experience challenges for designing future human-centered routine-aware systems?
\end{enumerate} 
These research questions were formulated to clarify the foundational structure of routine computing as a research area. \textbf{RQ1} aims to identify the underlying dimensions that organize existing approaches and to establish a shared conceptual vocabulary. At the same time, routine-aware systems are developed with diverse aims—from supporting health and aging to improving productivity or understanding community behavior—yet these motivations have not been systematically compared. \textbf{RQ2} addresses this by examining the functional goals that shape the development of routine-oriented technologies and the problems they aim to solve. Finally, as routine computing increasingly influences user-facing technologies, it raises questions about how such systems should interact with people, integrate into everyday life, and support long-term engagement. \textbf{RQ3} therefore explores the design and user-experience considerations that follow from routine-aware computation. Together, these questions provide a structured basis for synthesizing the literature and defining the scope of routine computing as an emerging research domain.

Therefore, in this paper, we conduct a systematic review of routine computing in HCI and ubiquitous computing, synthesizing 203 studies up to August 2025. Our analysis classifies work by sensing temporal granularity, behavioral and social contexts, cognitive and psychological factors, and adaptation strategies. The review examines routine-based applications in domains such as aging, health, self-reflection, and population-level insights. \\
Accordingly, this paper makes the following contributions:
\begin{enumerate}[nosep]
    \item We provide the first \textbf{systematic review} of routine computing in daily life, synthesizing 203 studies across sensing, modeling, and applied domains.  
    \item We articulate \textbf{routine computing as a conceptual foundation}, distinguishing it from related areas such as activity recognition and context-aware computing.  
    \item We develop a \textbf{taxonomy of routine computation} along temporal, behavioral, cognitive, and adaptive dimensions, consolidating diverse methods and approaches.  
    \item We identify \textbf{critical challenges and future opportunities}, including interpretability, personalization, ethical design, and data enrichment, setting a research agenda for advancing routine-aware HCI.  
\end{enumerate}

\section{Methodology}

\begin{figure*}[t]
    \centering
    \includegraphics[width=0.95\textwidth]{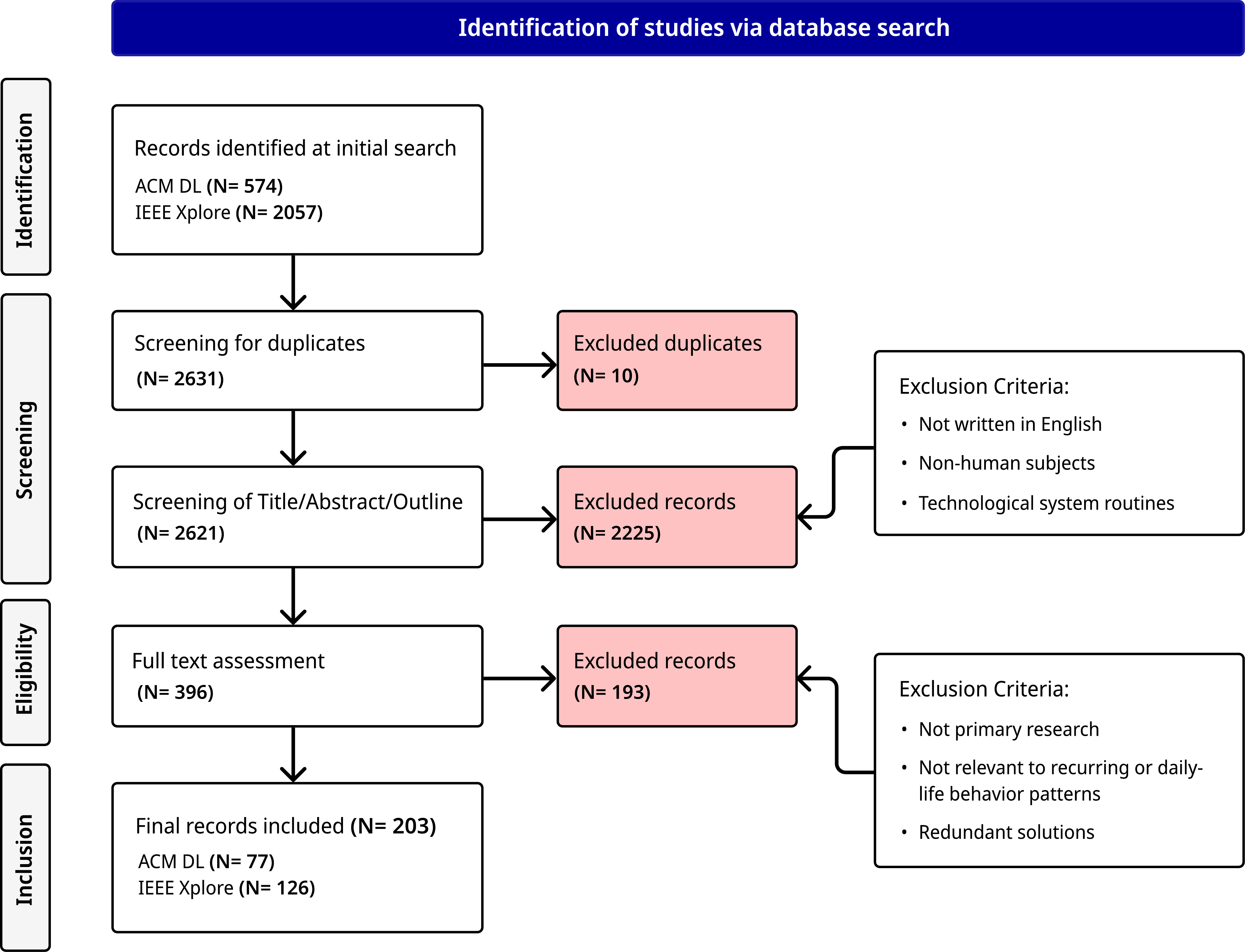}
    \caption{PRISMA flowchart detailing the article selection procedure}
    \Description{PRISMA flowchart detailing the article selection procedure. The initial search identified 2,631 records from two primary databases: ACM DL (574 records) and IEEE Xplore (2,057 records). After screening for duplicates, 10 were excluded, leaving 2,621 records for title, abstract, and outline review. During this screening, 2,225 records were excluded based on criteria such as not being written in English, non-human subjects, or focusing on technological system routines. A full-text assessment was then performed on the remaining 396 records. The final eligibility screen excluded 193 more records if they were not primary research, not relevant to recurring or daily-life behavior patterns, or presented redundant solutions. Ultimately, 203 records were included in the final review, with 77 papers from ACM DL and 126 from IEEE Xplore.
}
    \label{fig:flowchart}
\end{figure*}
\begin{figure*}[htbp]
    \centering
    \includegraphics[width=0.8\textwidth]{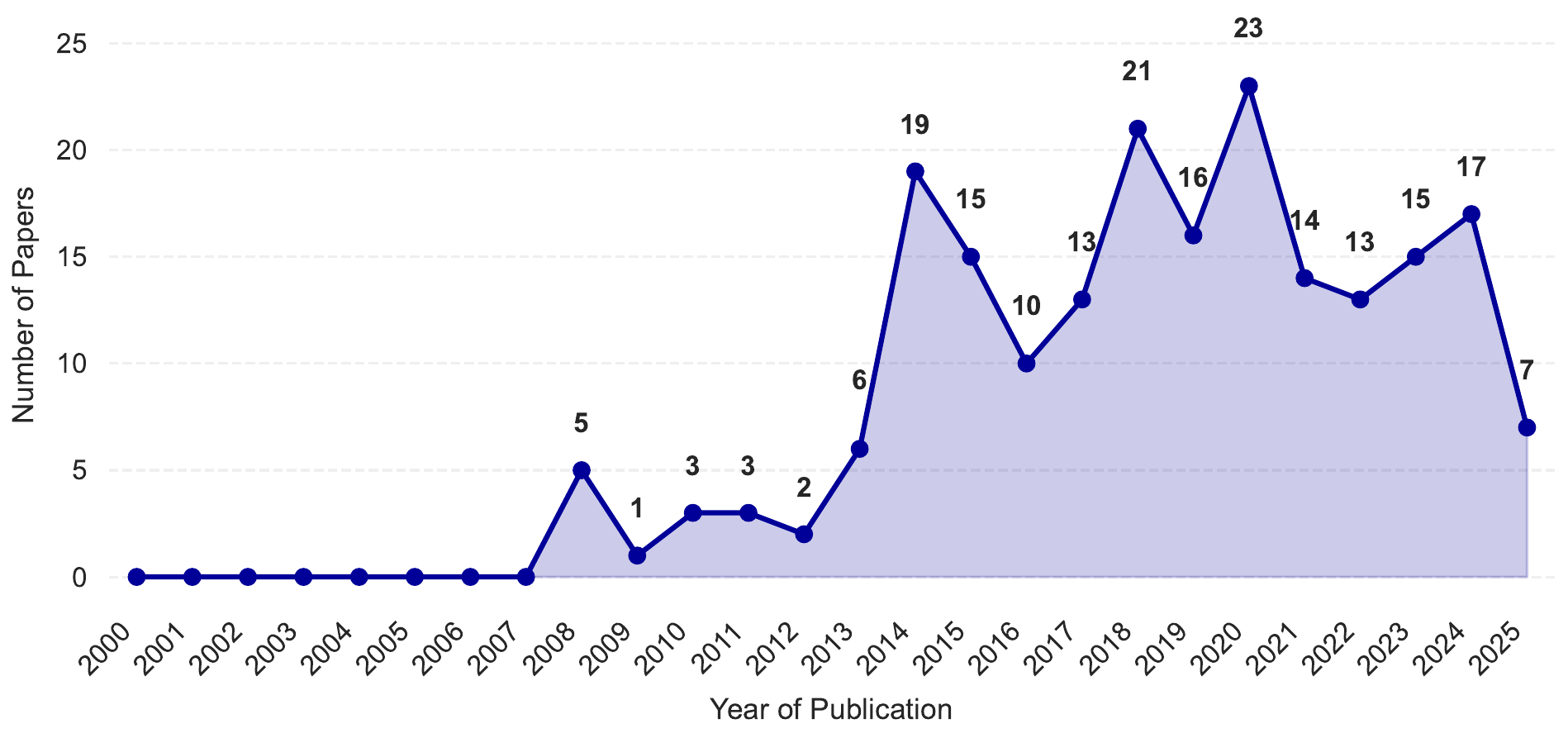}
    \caption{Annual distribution of included routine-computing publications that fulfill our criteria, illustrating the field's expansion since 2008. The chart only includes papers published up to August 31, 2025.}
    \Description{Annual distribution of included routine-computing publications that fulfill our criteria, illustrating the field’s expansion since
2008. The number of papers published in this field, as shown by the distribution, remained at zero or very low between 2000 and 2013. Significant growth began around 2014, when 6 papers were published, and peaked at 19 papers in 2015. The highest output was observed in 2020 with 23 papers. The research activity has remained relatively high since, with 21 papers in 2019, and continuing with 14, 13, 15, and 17 papers in 2021 through 2024 respectively. The chart only includes papers published up to August 31, 2025. At last, 7 papers were included 2025.}
    \label{fig:years-dist}
\end{figure*}
\subsection{Coding}
This review follows PRISMA 2020 guidelines \cite{page2021prisma}, ensuring a rigorous and transparent process for identifying and analyzing studies on routine computing in daily life. In accordance with the guidelines, this section consists of four key steps: 1) independent article screening by two researchers, 2) pilot annotation to identify key information for analysis, 3) iterative coding and establishment of a systematic codebook, and 4) synthesis of results and finalization of findings. 

\subsection{Research Strategy}

Initially, we conducted an exploratory survey of the routine-computing research landscape to identify the terminology most consistently used across relevant domains. We examined a total of 30 search results generated by different combinations of routine-related terms (e.g., “routine(s)", “daily life", and “ego-centric") together with behavioral and computational descriptors (e.g., “pattern", “behavior", and “modeling"), as well as domain-specific terms related to ubiquitous and context-aware computing. This iterative process ensured that our final keyword set aligned with how routine-oriented work is described across health, smart-home, mobility, and lifelogging research. For the final selection, we prioritized terminology appearing in foundational and frequently cited papers on ubiquitous computing, activity recognition, and emerging context-driven systems. Furthermore, the selection was supplemented by our established definition of the term “routine computing". Each group was combined with an “AND," yielding a total of 2,631 records (574 from ACM DL and 2,057 from IEEE Xplore). Next, a screening process was conducted to select papers within the survey's scope.

We restricted our search to the ACM Digital Library and IEEE Xplore, as our review focuses on practical, technology-oriented routine computation: systems that involve sensing, modeling, prediction, and empirical evaluation. These technical contributions are overwhelmingly published in HCI, ubiquitous computing, and pervasive sensing venues, which are concentrated within ACM and IEEE. Preliminary searches in SpringerLink, ScienceDirect, and PubMed/NIH showed that these databases primarily include biomedical, clinical, or theoretical work that does not involve computational routine modeling or deployable systems. We did not specify a time interval for the search, and the last search was conducted in August 2025.

\subsection{Screening}

We evaluated the relevance of all 2,631 papers returned by our search queries based on their titles and abstracts. Two authors independently screened all references in parallel and applied the following criteria for inclusion and exclusion: (1) we included only papers written in English; (2) we included only full papers published in peer-reviewed journals, conferences, symposiums, or workshops, thereby excluding technical reports, dissertations, work-in-progress papers, posters, patents, and tutorial or workshop proposals; (3) we filtered for topical match following our definition of routine computing and included papers that analyzed human behavior longitudinally, modeled recurring activity patterns, or used sensing technologies to characterize daily routines; (4) we excluded papers in which the term ``routine'' referred to technological or system-level processes such as network routines, workflow automation, cybersecurity routines, robotic task loops, or database maintenance; and (5) we excluded papers involving non-human subjects, including robotic, industrial, or animal behavior studies.

\begin{table}[htbp]
\centering
\caption{Final search keywords used in the literature review (combined with AND).}

\renewcommand{\arraystretch}{1.3}
\begin{tabular}{|p{0.95\columnwidth}|}
\hline
\large\textbf{Keywords (combined with AND)} \\[6pt]

\textbf{Routine-related terms}: 
``routine(s)'', ``daily life'', ``daily routine'', ``ego-centric'', ``proactive'' \\[4pt]

\textbf{Behavioral descriptors}: 
``activities'', ``pattern'', ``behavior'', ``user model(s)'' \\[4pt]

\textbf{Computational approaches}: 
``discover'', ``modeling'', ``recognition'', ``understand'', ``lifelog'', ``monitoring'', ``activity tracking'' \\[6pt]
\hline
\end{tabular}

\label{tab:search-keywords}
\end{table}

\subsection{Eligibility}

After title and abstract screening, 396 papers remained for full-text assessment. To determine eligibility for full inclusion after title and abstract screening, the same two authors independently reviewed all candidate papers at the full-text level, using the following additional criteria: (1) we included only papers providing clear methodological detail on sensing, modeling, or evaluating recurring human behaviors, excluding works that referenced routines without computational treatment; (2) we included only primary research contributions and excluded surveys, theoretical essays, opinion pieces, and purely conceptual frameworks; (3) we excluded papers that did not address recurring or temporally structured daily-life behavior, such as studies focused solely on algorithm optimization tasks, short-duration activities in a pre-set laboratory environment, or one-shot predictions without evaluation; (4) we excluded studies whose contributions were redundant with earlier work, for example follow-up papers reusing identical datasets and methods without offering substantial new insights.


Following this eligibility assessment, 203 papers (77 from ACM DL and 126 from IEEE Xplore) were retained for analysis. Inter-rater reliability between the two reviewers was evaluated using Cohen's~$\kappa$, which indicated substantial agreement ($\kappa = .833$, 95\% CI [.779, .888], $p < .001$). Any remaining disagreements between reviewers were discussed and resolved before final inclusion.

To systematically extract information from the corpus, we developed a structured data-extraction sheet consisting of 36 columns. The sheet was created through an initial open-coding phase in which both authors independently reviewed a subset of papers to identify recurring methodological and conceptual features relevant to routine computing. Codes were iteratively discussed and refined until consensus was reached.
The extraction schema captured several key aspects: (1) \textit{basic metadata} to contextualize each paper's scope and contribution; (2) \textit{sensing strategies and data-collection methods} to examine assumptions about how routines are captured; (3) \textit{modeled activities and computational techniques} to understand how routines are operationalized; (4) \textit{user-study characteristics and cognitive considerations} to reveal how human experiences and interpretations are incorporated; (5) \textit{evaluation metrics} used to assess performance, usability, or model quality; (6) \textit{the problem domain} to link routine-focused system goals to real-world applications; and (7) \textit{reported challenges} to identify recurring limitations across the field.
This structure ensured that the coding remained aligned with the research questions: \textbf{RQ1} focused on how routines are conceptualized and represented, while \textbf{RQ2} and \textbf{RQ3} guided attention to human-centered adaptation, usability, and evaluation practices. Importantly, by allowing patterns to emerge inductively from the corpus, this process enabled the identification of potential conceptual dimensions described in the next section.


\section[Dimensions of Human Routine Computation]{Dimensions of Human Routine Computation\footnote{In this section, $n$ refers to the total number of category instances identified across all reviewed papers, rather than the number of papers. (e.g., one paper can contribute more than one method category).}}
\label{sec:Section 3}
This section provides an integrated overview of human routine computation across 203 studies, structured around a set of core dimensions that emerged consistently throughout the corpus. The derivation of these dimensions followed an inductive–comparative process grounded in the coding framework developed during study extraction. Although the reviewed works vary in populations, sensing modalities, and analytical objectives, their approaches recurrently converged on several foundational features: the temporal organization of behavior, the types of activities and interactions being recognized, the cognitive or motivational processes underlying those behaviors, and the mechanisms through which systems adapt to change over time.

Across comparative analysis, these recurring features aligned robustly with four major analytical focuses that collectively reflect the conceptual pipeline of routine computing:
\begin{itemize}
    \item \textbf{Temporal granularity}—capturing when routines occur and how frequently they repeat, thereby revealing the rhythmic structure of everyday behavior.
    \item \textbf{Behavioral interaction context}—characterizing common activities and interactions with objects, bodies, and the surrounding environment.
    \item \textbf{Cognitive and motivational factors}—explaining the effect of integrating goals, intentions, and cognitive effort into routines.
    \item \textbf{Variability, adaptation, and breakdown detection}—describing how routines evolve, deviate, or reorganize in response to internal or external change.
\end{itemize}

The following sections examine each dimension in depth, outlining how it reflects a core dynamic of daily behavior—from temporal rhythms to contextual interactions.

\subsection{Temporal Granularity of Sensing Data (n=203)}
\label{sec:Section 3.1}
Capturing routines always requires attention to time, as the meaning of behaviors depends on their time organization and sequencing. We identified three levels of temporal granularity: longitudinal lifestyle analysis, multi-day routine analysis, and moment-to-moment actions, each reflecting different research goals, sensing strategies, and application contexts. A time distribution overview of the reviewed studies is presented in Table~\ref{tab:sensing_collection}.
\begin{table}[htbp]
\centering
\caption{Distribution of reviewed papers by collection time (six granular bins).}
\resizebox{\columnwidth}{!}{
\begin{tabular}{|p{0.26\columnwidth}|p{0.6\columnwidth}|r|}
\hline
\textbf{Collection time} & \textbf{Papers} & \textbf{Total (n)} \\
\hline
Seconds-Minutes & \cite{2,8,15,18,20,28,29,38,40,42,48,50,56,57,59,63,66,68,71,75,80,84,87,88,89,90,92,93,95,97,98,99,102,104,109,110,115,121,122,130,133,144,155,164,166,171,172,175, 190, 191, 195, 196, 197, 198} & 54 \\
\hline
Hour - Day & \cite{17,55,86,96,101,123,150,185,186,187,188} & 11 \\
\hline
Days - Week & \cite{24,44,51,78,85,100,105,113,117,134,159,160,165,184, 189, 192, 194, 199, 200} & 19 \\
\hline
Weeks - Month & \cite{11,12,14,23,30,34,35,36,37,43,45,46,47,49,52,53,62,64,69,76,79,82,106,107,112,118,124,125,126,135,138,139,140,148,152,157,177,181,182} & 39 \\
\hline
Months - Year & \cite{1,3,4,5,6,7,13,16,19,21,25,26,27,31,32,33,39,41,60,61,65,67,70,74,77,81,91,94,103,108,111,114,116,119,120,127,128,131,132,136,141,142,143,146,147,149,151,153,154,158,161,162,163,167,169,170,173,174,180,183,193, 201, 202, 203} & 65\\
\hline
More than 1 year & \cite{9,10,22,54,58,72,73,83,129,145,156,168,176,178,179} & 15 \\
\hline
\multicolumn{2}{|r|}{\textbf{Corpus total}} & \textbf{203} \\
\hline
\end{tabular}
}
\label{tab:sensing_collection}
\end{table}

\begin{figure*}[t]
    \centering
    \includegraphics[width=\textwidth]{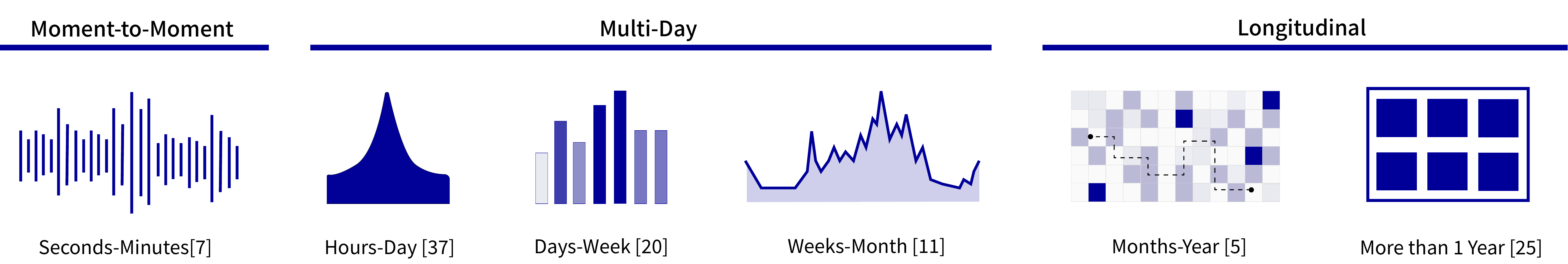}
    \caption{Instances demonstrating different temporal granularity patterns in sensing data.}
    \Description{ The figure contains a figure representing a basic structural or temporal component. 
There are several rhythms in the image: seconds-to-minutes actions, hours-to-day actions, days-to-week actions, and week-to-month actions, encapsulated under multi-day actions, and longitudinal actions of months-to-years or more than 1 year. Each factor has an abstract diagram representing it.
.}
    \label{fig:temporal}
\end{figure*}

\subsubsection{Longitudinal Lifestyle Analysis (n=80)}
Longitudinal studies reveal how routines develop over months or years, showing how repeated behaviors form lifestyle patterns. Low-burden, large-scale sensing is particularly effective for detecting gradual changes in health, energy use, or mobility that may be missed in daily observation. A group of studies $(n=18)$ focuses on long-term monitoring of sleep, stress, and mental states through continuous physiological monitoring \cite{163,178,179,193}. Another $(n=19)$ tracks household appliance usage and domestic routines via high-frequency sensing, including smart meters for electricity and gas, with environmental sensors capturing temperature, humidity, lighting, $CO_2$, sound, and pressure to contextualize activities such as showering \cite{31,67,103,156,77,202}. Mobility routines $(n=17)$ are monitored indoors and outdoors through GPS, cellular, BLE beacons, Wi-Fi CSI, PIR, and LiDAR, capturing movement patterns without labeling specific activities \cite{61,145,183,129,119,137,146}. Furthermore, 30\% of the studies $(n=26)$ analyze individual and population behaviors from large-scale digital traces such as social media, campus card use, location-driven patterns, and call logs to reveal urban mobility, consumption, and social routines \cite{21,131,147,174,201,203}.

\subsubsection{Multi-day Routine Analysis (n=69)}
To capture daily rhythms over days or weeks, this temporal perspective reveals how routines stabilize, adapt, or break down, shedding light on habit formation, daily structure, and rhythm deviations. Studies on personality and behavioral traits $(n=17)$ analyze multi-day patterns of device use, physical mobility, and social interaction to model stable tendencies with individual adaptation \cite{11,148,177,182,199}. Other work examines repeated routine action units such as meals, work, leisure, household tasks, exercises, and hygiene to uncover habitual structures and daily rhythms \cite{34,160,139,140,186,187,188, 192, 194}. Cognitive and affective daily patterns $(n=16)$ capture fluctuations in mood, stress, and attention through activity and behavioral data. They can detect sudden surges or spikes in emotional states, highlighting moments when routines are disrupted and enabling timely interventions in situations that require immediate attention or control \cite{12,138,134,53,52,200}. Research on habit formation and adjustment monitors recurring behaviors to identify emerging patterns and stability \cite{46,62,76}. Personalized self-reports {$(n=2)$} enhance people's ability to remember time-sensitive or health-critical activities \cite{184,189}. Finally, a major portion $(n=34)$ focuses on health and activity monitoring of elderly or vulnerable populations, such as visually impaired people, capturing temporal patterns in daily structure, physiological rhythms, and interactions with objects or appliances \cite{47,69,82,185}.

\subsubsection{Moment-to-Moment Actions (n=54)}
Momentary analyses focus on micro-scale human behaviors, where gestures, postures, and rapid adjustments unfolding in seconds or minutes reveal fine-grained patterns of routine. Hand and arm movements $(n=14)$ are tracked via wearable IMUs, smart rings, and gloves to capture gestures, grip adjustments, and object manipulations, enabling assessment of motor skills and task precision \cite{40,42,84,104}. Whole-body micro-movements $(n=9)$ are monitored using IMUs, pressure mats, and multi-sensor wearables, recording balance corrections, stance shifts, and micro-gait variations to evaluate stability and fatigue \cite{28,93,95,97,190,196}. Short, repetitive, or rhythmic actions $(n=9)$, such as typing, tapping, or exercise repetitions, are captured with high-frequency accelerometers and wrist-worn devices to assess temporal consistency and performance \cite{56,57,59}. Rapid locomotion and contextual micro-patterns $(n=22)$ examine moment-to-moment positional adjustments: some studies track short bursts of stepping, sidestepping, or gait corrections via IMUs and GPS \cite{8,18,20,80}, while others capture quick directional changes using environmental sensors, infrared sensors, or microphones with continuous event detection \cite{88,99,115,195}. Video-based surveillance focuses on abrupt body movements to identify dangerous shifts that precede falls \cite{191,197,198}.

\subsection{Behavioral Interaction Contexts (n=170)}
\label{sec:Section 3.2}
Routines are sustained through recurring behavioral interactions that span multiple levels of daily life, including immersion in the self, the body, the surrounding environment, and other people. These interactions unfold through habitual practices of personal health and wellness\cite{wood2007}. They are also expressed in embodied movements, shaped by the material and spatial advantages of technology and the surrounding environment \cite{gibson1979}. At the same time, they emerge as socially situated practices that derive meaning and stability from collective participation and cultural context \cite{shove2012dynamics}.
Together, these aspects highlight how routines are more than mere repetition but dynamic patterns that intertwine personal, material, and social layers of experience. Table~\ref{tab:behavioral_interactions} summarizes the core aspects identified in the reviewed studies, illustrating how researchers have operationalized and examined different behavioral layers of routine engagement.

\begin{figure*}[t]
    \centering
    \includegraphics[width=\textwidth]{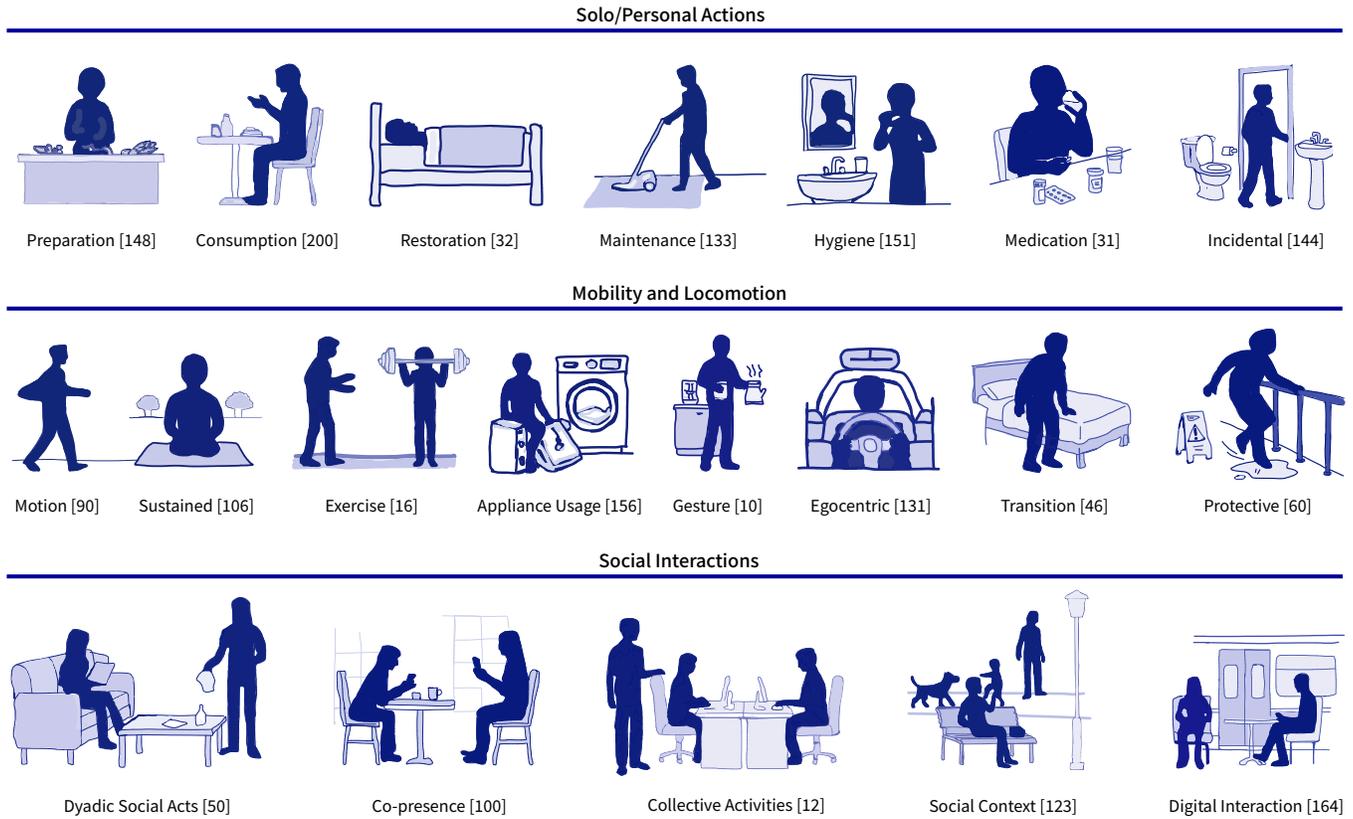}
    \caption{Instances demonstrating different behavioral interaction contexts.}
    \Description{The figure illustrates basic, routine behavioral activities of daily living and potential risks associated with mobility. 
There are 3 groups of activities: Solo/Personal Actions, Mobility and Locomotion Actions, and Social Interactions. The first group illustrates nourishing scenarios, sleep restoration, environmental maintenance, personal hygiene, medication management, and incidental self-care such as toileting. The second group shows other activities include motion, meditating, exercising, using appliances, egocentric activities(driving), and a critical protective concern: the risk of falling on a slippery surface. The last group illustrates social interactions, including dyadic social acts, group co-presence, collective social activities, social context within routines, and mediated or digital social interaction.
}
    \label{fig:behavioral}
\end{figure*}

\begin{table*}[htbp]
\centering
\caption{Summary of behavioral interaction subcategories in routine computing.}
\resizebox{\textwidth}{!}{
\begin{tabular}{|p{0.215\textwidth}|l|p{0.55\textwidth}|} 
\hline
\textbf{Main Category} & \textbf{Sub-Category} & \textbf{Paper IDs (Count)} \\
\hline
Solo / Personal Actions & Focused personal hygiene & \cite{7,15,25,26,34,62,75,82,83,106,117,133,150,159,163} (15) \\
 & Incidental and mandatory self-care & \cite{39,112,140,143,169,176} (6) \\
 & Nourishment preparation (IADL) & \cite{1,11,12,52,87,161,175} (7) \\
 & Nourishment consumption & \cite{73,172} (2) \\
 & Restorative sleep & \cite{4,5,6,9,10,16,19,33,51,134,152,162,164,167,168,179} (16) \\
 & Environmental maintenance (IADL) & \cite{35,36,46,68,105,144,149,156,165,173,182,192,197} (13) \\
 & Medication management (IADL) & \cite{49,53,57,70,74,91,102} (7) \\
\hline
Mobility and Locomotion & Ambulation and locomotion & \cite{3,20,28,29,63,66,80,93,96,130,177} (11) \\
 & Postural transitions & \cite{44,48,50,92,101,110} (5) \\
 & Sustained body positions & \cite{2,85,95,121,155} (6) \\
 & Exercise and deliberate movement & \cite{38,89,107,124,180,187} (6) \\
 & Gestures and object interaction & \cite{8,13,14,17,21,30,32,41,47,54,119,129,153,18,55,71,78,79,84,88,90,98,100,104,122,132} (27) \\
 & Appliance and device-centric & \cite{23,27,31,60,64,67,103} (7)\\
 & Anomalous and protective movements & \cite{37,40,97,109,198} (5) \\ 
 & Egocentric motion view & \cite{12,22,189,198} (4) \\ 
\hline
Social Interactions & Dyadic social acts & \cite{22,56,111,114, 136,189} (6) \\
 & Group co-presence and co-location & \cite{69,81, 151,201, 203} (5) \\
 & Collective social activities & \cite{61,108,113,145,146, 147,191} (7) \\
 & Social context within routines & \cite{58,72,118,125, 131} (5) \\
 & Mediated or digital social interaction & \cite{126,135,148,24,43,65,77,115,123,142} (10) \\
\hline
\end{tabular}
}
\label{tab:behavioral_interactions}
\end{table*}

\subsubsection{Self-Care Interactions (n=66).}

Research on habits indicates that self-care routines often run on “autopilot", with hygiene, nourishment, and sleep reducing cognitive effort for complex tasks \cite{wood2007}. Focused personal hygiene $(n=15)$ includes showering, handwashing, and brushing teeth, supporting health and social readiness \cite{62,75,82,83,159}. Incidental self-care $(n=6)$ covers need-driven actions such as toileting and hydration, where frequent and unpredictable interruptions shape daily patterns \cite{140,143,169,176}. Nutrition management $(n=7)$ involves food preparation and cleanup, where deviations reveal cognitive, social, or motivational changes \cite{1,11,12,87,161}, while nourishment consumption $(n=2)$ captures eating and drinking patterns that structure routines and foster social participation \cite{73,172}. Restorative sleep $(n=16)$ reflects the quality of habitual rest, which strongly influences physical and cognitive function \cite{33,152,168,179}. Environmental maintenance $(n=13)$, such as cleaning, vacuuming, and organizing of living space, signals physical or motivational capacity \cite{35,165,173,192}. Medication management $(n=7)$ emphasizes the adherence to rule-based and contextual reasoning based on observation of events, where timing is critical for health outcomes \cite{53,57,74,91,102, 197}.

\subsubsection{Physical \& Mobility Interactions (n=71).}
From an ecological perspective, mobility is more than movement through space. It reflects how people act on environmental opportunities, revealing both capabilities and constraints \cite{gibson1979}. Ambulation and locomotion $(n=11)$ capture goal-directed movements such as walking, running, or climbing stairs, where path and gait indicate ability and fatigue \cite{80,93,96,130,177}. Postural transitions $(n=5)$, such as getting out of bed, highlight how speed and stability shape the initiation of activities \cite{44,50,101,110}. Sustained body positions $(n=6)$ reflect comfort, engagement, or health risks by location and duration of posture \cite{2,85,95,121,155}. Exercise and deliberate movement $(n=6)$ cover structured activities such as workouts, yoga, or rehabilitation, where consistency and form reveal motivation and intensity, indicating whether individuals promote their health \cite{38,89,107,124,180,187}. Object manipulation $(n=27)$ involves handling items—pouring, stirring, writing—that combine into larger tasks like cooking or drinking \cite{47,55,100}. Appliance and device-centered interactions $(n=7)$ include daily operations of lights, stoves, or kettles \cite{64,67,103}. Anomalous and protective movements $(n=5)$, such as falls or bracing, signal immediate safety risks \cite{37,40,97,109,198}. Finally, egocentric recording $(n=4)$ captures fine-grained body movements and attention shifts that are often missed by stationary or device-centric sensors \cite{12,22,189,198}.

\subsubsection{Social Interactions (n=33).}

Social practices are fundamental to routines, as daily life is organized not only by individual habits but also by shared norms, roles, and collective activities \cite{reckwitz2002, pentland2005}. Dyadic social acts capture one-to-one encounters, such as handshakes, greetings, and phone calls $(n=6)$, reflecting direct physical or verbal exchanges between two people \cite{111,114,136,189}.
Group co-presence and co-location $(n=5)$ emphasize shared spatial proximity, such as class or workplace attendance inferred from Bluetooth or mobility data \cite{69,81,151,201,203}.
Collective social activities $(n=7)$ involve coordinated participation in leisure or community events, including shopping, parties, concerts, dining, and therapy sessions, or opportunistic encounters based on shared interests \cite{61,146,147,191}.
Social context within routines $(n=5)$ highlights how the presence of partners, family, or colleagues shapes choices of daily activity and role-based behaviors \cite{118,125,131}.
Mediated or digital social interaction $(n=10)$ involves technology-based exchanges such as messaging, planning, and documentation mediated by devices \cite{24,148}.

\subsection{Cognitive and Psychological Factors (n=83)}

\label{sec:Section 3.3}
Routines are shaped by cognitive and psychological processes that influence how people remember, sustain, and reflect on their actions. We identify five factors: prospective memory, intention and motivation, reward sensitivity, cognitive effort, and self-reflection, which together explain how routines connect to memory, meaning, and personal values. An overview of these factors is summarized in Table~\ref{tab:cognitive_psychology}.

\subsubsection{Remembering Future Tasks (Prospective Memory) (n=4).}
Prospective memory theory emphasizes how systems support users in keeping future intentions in mind, such as taking medication or attending scheduled meetings\cite{mcdaniel2000}. For instance, one study provides reminders for deviations from expected routines \cite{136}, while another predicts users' upcoming actions in real time to anticipate intended behaviors \cite{1,181}. Moreover, prompting users to complete self-reports enhances accurate recall of daily activities \cite{184}. These approaches highlight the importance of external aids in helping users remember what they plan to do, rather than relying solely on immediate cues.

\begin{figure*}[t]
    \centering
    \includegraphics[width=\textwidth]{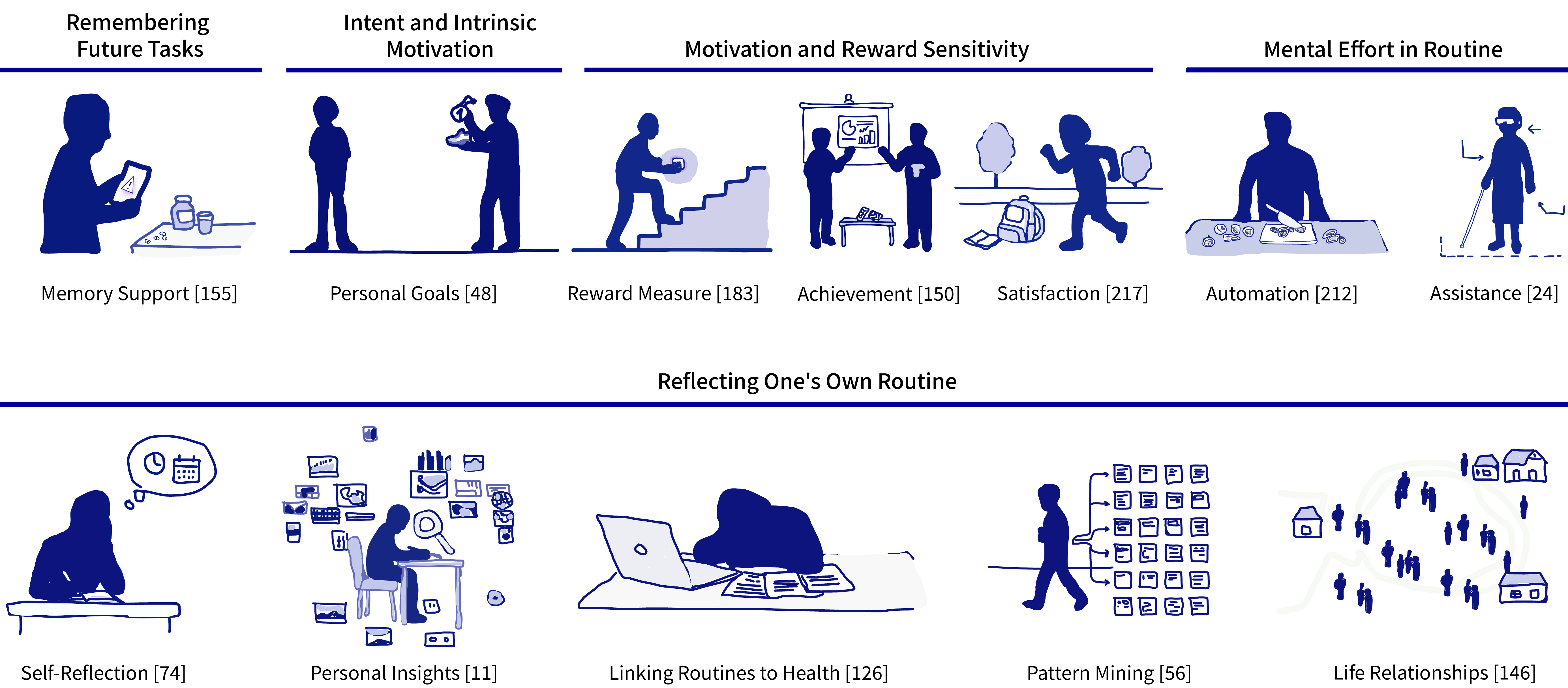}
    \caption{Instances demonstrating different cognitive and psychological factors. }
    \Description{The figures illustrate various complex cognitive functions and interactions with information and the environment. 
There are several factors involved: remembering future tasks, following personalized goals, measuring rewards, accomplishing achievements, having satisfaction, relying on automation, and utilizing technological assistance for visually impaired situations. Another line of the image includes illustrations regarding self-reflection, personal insights, routine demonstration by reference to personal health, such as sleep patterns, pattern mining, and analyzing life relationships. All images are abstract representations of those cognitive functions. Activities involve understanding warnings or contraindications, particularly with medication. Other figures show complex actions and professional tasks like presenting information. The illustrations also cover self-reflection functions, such as time/date planning, information processing and recall, and navigating and managing a high volume of documents or data.
}
    \label{fig:cognitive}
\end{figure*}

\begin{table*}[htbp]
\centering
\caption{Summary of cognitive and psychological subcategories in routine computing.}
\resizebox{\textwidth}{!}{
\begin{tabular}{|l|l|p{5cm}|}
\hline
\textbf{Main Category} & \textbf{Sub-Category} & \textbf{Paper IDs (Count)} \\
\hline
Remembering Future Tasks & Prospective memory support & \cite{1,136,181,184} (4) \\
\hline
Intent and Intrinsic Motivation & Linking actions to personal goals & \cite{124,139,180,182} (4) \\
\hline
Motivation and Reward Sensitivity & Measuring affective rewards & \cite{26,117,151,200} (4) \\
 & Leveraging reward sensitivity for behavior change & \cite{172,178} (2) \\
 & Satisfaction, achievement, and pleasure in routines & \cite{147,148,174} (3) \\
\hline
Mental Effort in Routines & Automating complex activities & \cite{17,25,29,51,72,82,83,85,89,93,94,103,105,106,111,113,145} (17) \\
 & Reducing cognitive load through context inference & \cite{7,100,120,122,133,141,144,160,189} (9) \\
\hline
Reflecting on One's Own Routines & Self-reflection feedback loop & \cite{6,13,16,30,49,64,67,112,129,154,156,176,179} (13) \\
 & Modeling personal data for self-insight & \cite{34,43,69,115,118,134,135,162, 171, 191} (10) \\
 & Pattern mining and routine discovery & \cite{24,77,81,201,203} (5) \\
 & Linking routines to health and cognitive states &  \cite{39,53,163,164,168,185}  (6) \\
 & Quantifying life rhythms and causal relationships & \cite{125,137,152,161,165,177} (6) \\
\hline
\end{tabular}
}
\label{tab:cognitive_psychology}
\end{table*}

\subsubsection{Intent and Intrinsic Motivation (n=4).}
Self-determination theory emphasizes that routines sustained by intrinsic motivation linked to autonomy, purpose, or enjoyment are more resilient than those driven solely by external demands, highlighting the need for systems that align with personal goals and values\cite{deci1985}. Systems that generate personalized “if-then" plans link situational cues to desired behaviors, strengthening intention-driven actions \cite{124}. Others examine intrinsic motivations for eating \cite{139} or exercising \cite{180}, revealing how pleasure, health, and personal goals drive adherence. Research also highlights how feelings of competence or incompetence influence participation, showing the psychological difference between a task users feel compelled to do versus one they truly want to pursue \cite{182}.
\subsubsection{Motivation and Reward Sensitivity (n=9).}
Research in behavioral psychology\cite{skinner1953} shows that routines are strengthened when actions are paired with rewarding outcomes, whether through pleasure, achievement, or relief, underscoring how technologies can harness reinforcement to maintain engagement. Measurement of emotional states, stress, and arousal \cite{26,117,151,200} exemplifies how routines align with positive or negative outcomes as well as the duration and intensity of the experience. Research \cite{147,148,174} highlights how social influences and individual personality traits affect the satisfaction and achievement of daily activities. Others leverage reward sensitivity to support behavior change, such as smoking cessation \cite{172} or interventions for mental well-being \cite{178}, showing how systems can strategically harness motivational levers to maintain engagement.
\subsubsection{Mental Effort in Routines (Working Memory and Cognitive Load) (n=26).} Cognitive load theory suggests that routines falter when tasks demand too much memory or attention, positioning automation and context inference as crucial ways to ease effort and support sustainable behaviors \cite{sweller1988}. Systems that automate activity recognition \cite{17,29,83}, segment complex action sequences \cite{89,94,122}, or infer hierarchical context \cite{120,133,141,144} reduce cognitive load, allowing users to execute routines without being overwhelmed. Some focus on eye-context analysis to reveal which routine steps strain visual attention.\cite{189} Others simplify repetitive or attention-demanding tasks, such as automatically adjusting smartphone settings \cite{100} or detecting object interactions \cite{160}, freeing working memory to support smoother and more manageable routines.
\subsubsection{Reflecting on One's Own Routines (Metacognition and Self-Monitoring) (n=40).}
\label{sec:Section 3.3.5}
Feedback from self-monitoring tools that illustrate vital patterns enables adaptive behavior and often changes people's routines, as metacognitive research \cite{flavell1979} shows. Approximately 63\% of the cognitive studies highlight how systems enable users to notice and rethink their habits. Some focus on providing a feedback loop for self-reflection \cite{6,13,16,30,49}. Others collect and model personal data, such as movement, smartphone usage, or digital footprints, to provide users with insights about their habits \cite{69,115,134,191}. Systems that mine patterns \cite{24,77,81} or expose location-dependent bottlenecks\cite{201,203} in daily tasks help users pinpoint exactly where their routines break down and why. Linking routines to health outcomes (e.g., stress, cognitive decline, sleep quality, and aerobic activity) allows users to understand the impact of manner on well-being \cite{39,83,163,168,185}. Advanced approaches quantify life rhythms \cite{161,152} or model causal relationships between context and action \cite{137}, supporting deeper metacognitive awareness and enabling informed behavior change.

\begin{figure*}[htbp]
    \centering
    \includegraphics[width=\textwidth]{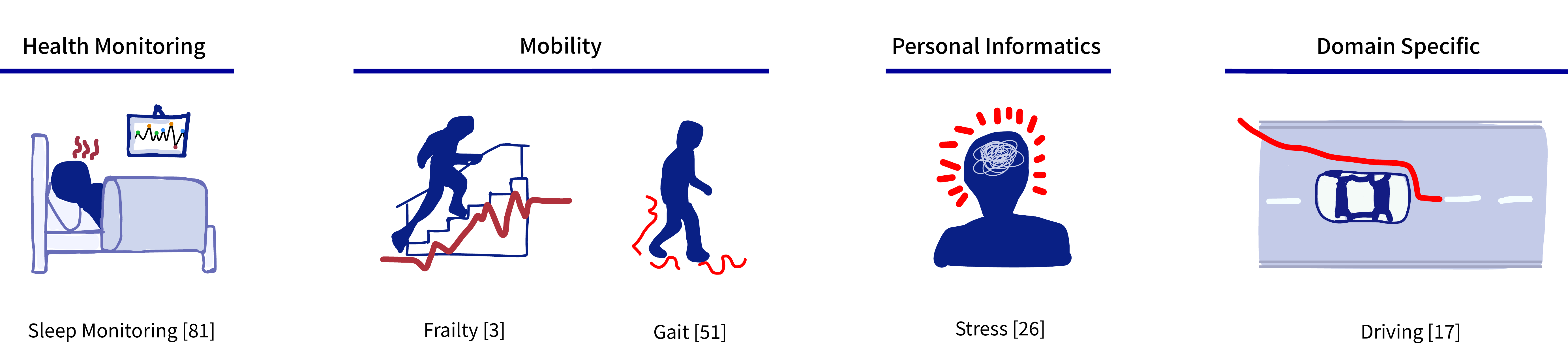}
    \caption{Instances demonstrating different variability, adaptation, and breakdown detection in routines. }
    \Description{This figure illustrates challenges associated with variability, adaptation, and breakdown detection in routines. The first icon shows challenges in safely climbing stairs, referencing impaired motor control under stress or cognitive strain. The second image displays a person with unsteady gait, indicating difficulty adapting motor actions. The third image depicts a head with tangled lines and outward stress marks, symbolizing overload, confusion, or disorientation. Another shows a car veering off the expected path, representing failure in planning or attention during complex tasks such as driving. }
    \label{fig:variability}
\end{figure*}

\subsection{Variability, Adaptation, and Breakdown Detection in Routines (n=83)}
\label{sec:Section 3.4}

This dimension considers the factors driving routine deviations, along with intervention strategies for breakdowns and models of adaptation over time. An overview of the reviewed studies is summarized in Table~\ref{tab:app-domain-categorization}.

\subsubsection{Core variables of routine deviation criteria} 
The highlighted variables that structure routines are the specific temporal, contextual, and behavioral dimensions. Common distinctions include weekday vs. weekend schedules, activity ordering, and contextual modifiers such as weather, area of interest, or social presence \cite{10,30,108,137,192}. Others emphasize performance-based indicators such as frailty, stress levels, or sleep patterns as critical variables for understanding deviations \cite{6,117,167,168}. Such factors are often incorporated into models to capture the fluidity of routines and inform adaptive systems. Some works emphasize domain-specific criteria, such as falls detected through accelerometer thresholds or abnormal nutritional expenditure in student routines \cite{93,109,73}. 

\subsubsection{Intervention strategies for routine support}
\label{sec:Section 3.4.2}
During the anomaly detection process, the systems are set to act accordingly in case of an emergency. Many provide alerts or notifications to caregivers, family, or clinicians \cite{10,70,170,179}. Others intervene directly through reminders, feedback, or personalized recommendations to nudge user behavior \cite{34,124,168,178}. A smaller group explores novel interaction strategies, such as assistive robots for confirmation feedback \cite{16} or visualization tools to reframe disruptive behavior \cite{177}. 
\subsubsection{Principles for long-term model adaptation} Addresses how systems evolve alongside changing routines. Techniques include dynamic clustering and feedback loops to incorporate new data \cite{1,16,30}, periodic pattern mining that adapts to contextual shifts, such as seasons or semesters \cite{74,145}, and semi-supervised retraining approaches to handle gradual change \cite{49,165}. Using sliding-window multi-HMMs, the model adapts to variable routines, helping people with chronic illnesses to track medication and self-care tasks even on highly irregular days \cite{193, 201, 202}. Other models explicitly incorporate user input or contextual queries, enabling digital twins to adapt flexibly to changes \cite{111,129}. 

\begin{table}[htbp]
\centering
\renewcommand{\arraystretch}{1.5} 
\caption{Categorization of anomaly detection papers by primary application domain.}
\resizebox{\columnwidth}{!}{
\begin{tabular}{|p{0.35\columnwidth}|p{0.75\columnwidth}|}
\hline
\textbf{Application Domain} & \textbf{Paper IDs and Key Characteristics} \\
\hline
\textbf{Health Monitoring} & \textbf{Papers:} \cite{1,6,9,10,16,30,39,41,47,51,53,59,61,67,70,83,91,111,129,136,159,167,169,170,176,179, 184, 185, 187, 192, 193, 197, 198} (33 papers) \\
& \textbf{Key Variables:} Room transitions, appliance usage, sleep patterns, medication intake, and frailty indexes. \\
& \textbf{Anomalies:} Missed activities, unusual inactivity, deviations from long-term routines. \\
\hline
\textbf{Mobility } & \textbf{Papers:} \cite{2,31,34,46,49,64,65,74,103,108,124,145,149,165,181,40,93,97,109,110,185,  186,187,194,203} (25 papers) \\
& \textbf{Key Variables:} Acceleration, posture, location, pressure, WiFi CSI signals. \\
& \textbf{Anomalies:} Falls, unstable gait, abnormal transitions (e.g., sit-to-stand). \\
\hline
\textbf{Personal Informatics} & \textbf{Papers:}  \cite{11,12,14,43,73,117,137,152,168,174,178,180,182,186,188,189,191,195,199,200,201,203} (22 papers) \\
& \textbf{Key Variables:} Sleep stages, physiological signals (HR, EDA), meal patterns, exercise, screen time. \\
& \textbf{Anomalies:} Poor sleep quality, high stress, unhealthy eating, sedentary behavior, and goal failure. \\
\hline
\textbf{Domain-Specific} & \textbf{Papers:} \cite{84} (Handwriting Analysis), \cite{171} (Youth Behavior), \cite{177} (Driving Behavior) (3 papers) \\
& \textbf{Key Variables:} Pen gestures, step counts, driving maneuvers. \\
& \textbf{Anomalies:} Tremors, unusual student activity, aggressive driving. \\
\hline
\end{tabular}%
}
\label{tab:app-domain-categorization}
\end{table}
\subsection{Conceptual Framework for Routine Computing}

\begin{figure}[htbp]
    \centering
    \resizebox{\columnwidth}{!}{%
    \includegraphics{section/figure/framework.pdf}
    }
    \caption{Conceptual framework diagram showing interdependent Context, Cognition, and Adaptation dimensions within the overarching layer of Temporal Rhythm, which enables routine computation.}
    \Description{Conceptual framework model structured as a Venn diagram, illustrating the intersection of three fundamental routine computing dimensions: Adaptation, Context, and Cognition within the overarching layer of the Temporal Rhythm dimension, which enables routine computation.}
    \label{fig:framework}
\end{figure}

Figure~\ref{fig:framework} illustrates the three core dimensions---\textit{Context}, \textit{Cognition}, and \textit{Adaptation}---bounded by the outer layer of \textit{Temporal Rhythm}, which forms the conceptual foundation of routine computing. 

The outer, encapsulating role of \textit{Temporal Rhythm (\ref{sec:Section 3.1})} is intentional because time and temporal structure are the practical foundation of routine computing: every system in our corpus depends on sensing streams (temporal traces of behavior, location, physiology, and device logs) to reveal repetition, periodicity, and change. Framing temporal rhythm as the prerequisite layer highlights sensing as the data substrate of routine intelligence and explains why temporal granularity determines the feasibility of analyses and interventions \cite{dey1999,markweiser1999}. 

The linkage between \textit{Context (\ref{sec:Section 3.2})} and \textit{Cognition (\ref{sec:Section 3.3})} is shown as interdependent because cognitive models of routine behavior do not operate in a vacuum—interpretation, intention, and memory are shaped by the environment and the available contextual cues. Accurate mental models (prospective memory, goals, workload) require context to ground hypotheses about what an actor can or will do. This idea is supported by work on embodied and distributed cognition and by classical theories of affordances \cite{Hutchins1995-HUTCIT, Clark1981-CLABTP}. Showing cognition nested with and tied to context thus makes explicit that sensing, situational cues, and mental modeling are a single epistemic chain.

Finally, \textit{Adaptation (\ref{sec:Section 3.4})} is drawn as bidirectional and permeable because adaptive interventions both depend on—and actively alter—contextual inputs and cognitive states. Changes produced by adaptive systems (reminders, rescheduling, nudges, personalized feedback) can modify how often behaviors occur, which contextual cues are salient, and how users internally represent tasks or goals; conversely, shifts in context or cognition require new adaptation \cite{wood2007, oinas2010}. Representing adaptation as a cross-cutting, change-producing mechanism highlights its dual role as an analytic target and as an instrument for influencing the other dimensions, aligning with empirical and theoretical work on habit dynamics and behavior-change design. 
Innovation often arises at the intersections of these dimensions, enabling researchers to define novel design questions that go beyond single-focus solutions. For instance, analyzing the \textit{Context} $\cap$ \textit{Cognition} relationship can inspire models that use rich sensing to predict human intention within a routine, while the \textit{Cognition} $\cap$ \textit{Adaptation} link can guide the design of interventions whose intrusiveness is dynamically tuned by the system's prediction certainty.

By making these hierarchical and functional relationships explicit, the framework serves as a practical mental model to position future projects within the field, supporting clearer comparison across studies and guiding researchers in designing comprehensive systems that seamlessly integrate sensing, modeling, understanding, and intervention to advance routine-aware technologies.

\section{Goals of Computing Human Routines in Daily Life}
\label{sec:Section 4}

Our review of the literature identified diverse goals for routine computing in daily life, reflecting the different populations, contexts, and scales addressed. Across the studies, we identified four primary goals: supporting aging and disability care, promoting healthy habit formation, enhancing self-reflection through intelligent assistance, and informing community-level understanding through aggregated routines. These goals shaped the behaviors researchers chose to model, the sensors and data modalities emphasized, and the design of supportive technologies. Also, they illustrate how real-world applications extend and operationalize the routine computing dimensions synthesized in Section \ref{sec:Section 3}, grounding those abstractions in concrete use contexts. Moreover, we analyzed how the pursuit of each goal positioned routine computing at different levels of impact, ranging from individual well-being to collective social development.

\subsection{Supporting Routines in Aging and Disability Care (n=46)}
\label{sec:Section 4.1}

Monitoring human routines in the context of aging and disability care serves two complementary purposes: while disability-focused studies emphasize health-related pattern recognition to support diagnosis and rehabilitation, aging-focused work highlights the tracking of daily routines and risks to promote independence and timely caregiving support.

\subsubsection{Monitoring Health-Related Patterns in Disability Contexts (n=11)}

In disability care, growing research emphasizes the use of routine-oriented sensing to capture daily behaviors as proxies for functional ability, disease trajectories, and rehabilitation progress. A variety of wearable and object-embedded systems have been introduced to track physical activity patterns and movement characteristics, supporting the assessment of motor impairments and neurological conditions \cite{29,92,104,122,134}. Likewise, ambient and smart home infrastructures integrate multimodal sensing—from cameras and appliance usage to localization technologies—to monitor activities of daily living, detect cognitive decline, and anticipate risks in dementia or physical disability care \cite{39,60,67,112,193}. Recent studies also indicate that sequence learning and predictive modeling have advantages in recognizing temporal dependencies in patients' daily behaviors, thereby improving the accuracy of activity recognition and the ability to forecast deviations in routine that may signal emerging health concerns \cite{45,60,112}.

\subsubsection{Tracking Daily Routines and Risks in Aging (n=35)}

%
A central goal of monitoring routines in aging contexts is to support older adults in maintaining independence while enabling caregivers and family members to understand their daily lives better. Routine-oriented sensing has been used to capture mobility, exercise adherence, and everyday activities, thereby offering valuable insights into functional status, early frailty, and lifestyle changes that may require intervention \cite{1,3,6,9,44,47,74,82,93,110,140,155,179}. Video surveillance approaches allow for non-intrusive monitoring for risk scenarios such as falls or wandering \cite{196,197,198}. These insights are particularly valuable in community-dwelling contexts, where sustained monitoring captures gradual behavioral shifts that might otherwise go unnoticed, offering an early window into emerging health concerns \cite{36,61}. Beyond these settings, routine-oriented systems also support broader caregiving with continuous awareness of older adults' daily well-being and potential care needs \cite{10,70,91,170,176,192}. A particularly promising line of work builds on digital twin concepts, integrating comprehensive data streams to construct dynamic representations of older adults' daily routines, allowing more holistic and proactive monitoring \cite{33,111,129,190}.

\subsection{Promoting Everyday Health Practices and Specific Habit Patterns (n=55)}


Everyday health routines encompass both broad patterns, such as mental well-being and the balance between physical activity and work, as well as more specific lifestyle habits. Together, these routines offer valuable perspectives for understanding and supporting everyday health management.

\subsubsection{Indicators of Mental Health States (n=10)}

Daily routines provide a valuable lens for understanding mental health, as subtle variations in activities, mobility, and physiological signals often reflect underlying emotional states. Studies have shown that wearable and smartphone-based sensing can infer mood, stress, and affective responses from multimodal data such as activity patterns, sleep, diet, and biometrics \cite{7,26,117}, while longitudinal monitoring further enables the detection of routine changes that may indicate depression or heightened stress levels \cite{49,126,163}. Integrating heterogeneous data streams has been found to improve the assessment of mental well-being, offering richer and more reliable indicators than single-source approaches \cite{26,114,163,174,193}. Also, frameworks have begun linking routine-aware detection to actionable recommendations, highlighting the potential of computing human routines to support proactive strategies for maintaining and enhancing mental health \cite{7,178}.

\subsubsection{Equilibrium between Physical Activity and Work (n=31)}

Balancing physical activity and work-related routines is a key objective in understanding human daily practices. Studies have emphasized the importance of recognizing locomotion, gait, exercise behaviors, and occupational tasks as interconnected elements of health and productivity \cite{28,35,37,38,40,66,85,187}. In contrast, others highlighted how daily schedules reveal interdependencies between mobility, work patterns, and self-aware activities \cite{32,54,58,65,186}. In addition, smartphone and multimodal sensing extend this perspective by situating physical activity within broader lifestyle demands, including maintaining exercise plans, balancing work responsibilities with periods of rest, and engaging in leisure or social activities \cite{69,76,105,120,195,199}. These insights point to proactive support systems that help individuals maintain an equilibrium between occupational requirements and health-promoting routines, offering opportunities for both performance enhancement and long-term well-being \cite{86,89,124,132,180,200}.

\subsubsection{Fine-Grained Lifestyle Patterns for Health Promotion (n=14)}

Understanding lifestyle at a fine-grained level is critical for promoting sustainable health behaviors. Studies have demonstrated the value of capturing specific routines such as eating \cite{12,139,161}, sleep \cite{152,161,168}, smoking \cite{172}, hygiene \cite{15,143}, household tasks \cite{202,188}, and micromotor activities such as handwriting \cite{84} or scratching behaviors \cite{150}. These detailed patterns provide insights into the subtle, often-overlooked aspects of daily life that actually sometimes affect long-term well-being. Complementary approaches utilize smartphones and wearable devices as personal informatics tools to track multidimensional routines and support behavior change through personalized feedback and goal-setting \cite{34,127}. Together, these works highlight how routine-oriented sensing can reveal the causal links between small-scale habits and broader health outcomes, therefore enabling more targeted interventions and fostering individual awareness and self-regulation \cite{152,161,168}.

\subsection{Designing Adaptive Support through Reflection, Assistance, and Prediction of Human Routines (n=49)}


Computing human routines not only enables individuals to reflect on their everyday behaviors but also supports adaptive guidance through context-aware systems. Moreover, it facilitates forward-looking strategies, such as routine prediction and synthetic data generation. These directions illustrate a progression from personal self-awareness toward proactive and assistive technologies.

\subsubsection{Fostering Self-Reflection in Living Routines (n=30)}


Self-reflection in daily life refers to the capacity of individuals to observe and evaluate their own routines, thereby recognizing healthy patterns, imbalances, or disruptions that may affect well-being. In smart home contexts, making routine structures visible, such as meal times, mobility behaviors, electricity usage, or leisure practices, can help individuals better understand their habits and foster healthier lifestyle management \cite{4,5,25,52}. Meanwhile, systems that highlight uniformity or deviations in routines allow users to reflect on adherence and recognize potential risks or lifestyle drifts \cite{13,64,77}. Personalized self-reports are triggered based on optimized sampling timeframes to help users reflect on their routines without causing disruption \cite{154,184}. Beyond domestic settings, workplace-oriented sensing approaches similarly encourage awareness of daily rhythms and vitality \cite{115}. Another strand of research emphasizes interpretability, for example, by representing activities through human-object interactions or semantically structured action sequences, which makes daily behaviors easier to recall, compare, and reason about in ways that support self-awareness \cite{27,55,78,79,90}.

\subsubsection{Guiding Human Behaviors with Context-Aware Assistance (n=14)}
\label{sec:Section 4.3.2}

Context-aware assistance aims to guide people in maintaining healthy and productive routines by adapting system responses to ongoing behaviors. Such systems often model the probabilistic sequencing of activities and detect deviations that may indicate risks, allowing timely interventions \cite{11,16,41}. Assistance can take the form of alternative recommendations or adaptive feedback loops, where robots, mobile devices, or ambient systems communicate with users and incorporate their input to refine subsequent guidance \cite{16,43,53,106}. More importantly, interventions are not limited to health and safety but also support cultural or religious practices, productivity, and consumption, thereby embedding technology into broader aspects of daily life \cite{20,113,118,142,157}. Moving beyond passive monitoring to proactive support, these systems exemplify how routine computing can enhance well-being and autonomy through contextually informed assistance \cite{43,53,113,191}. Visually impaired people, for example, benefit from context-aware navigation aids that adapt to their routines and environments, improving mobility and independence \cite{185,189}.

\subsubsection{Advancing Routine Prediction and Synthetic Data Generation (n=5)}


Routine prediction and synthetic data generation have emerged as important research directions in routine modeling. Recent work develops models that capture spatiotemporal patterns of activities, enabling the clustering of ADL sequences and the creation of synthetic variations to test system robustness \cite{19}. Predictive frameworks using motion sensors or IoT devices further support anomaly detection and daily behavior forecasting without relying heavily on manual labeling \cite{30,116,177}. Other studies, such as collaborative intent prediction, leverage contextual signals to anticipate real-time user needs, extending routine modeling into digital spheres\cite{181}. Such advances explore how prediction and synthesis can inform human-centered design and research toward adaptive systems.

\subsection{From Individual Routines to Population-Informed Insights (n=9)}
\label{sec:Section 4.4}

Population-level analyses of human routines extend the scope of routine computing beyond individual assistance, offering broader perspectives on mobility, social interaction, and collective behavior. By examining aggregated digital traces and behavioral datasets, these studies illuminate how everyday activities scale to inform domains such as public health, education, and smart city management.
Large-scale behavioral mining has revealed recurring mobility and lifestyle patterns in urban and community settings, showing how periodic routines can inform city planning and management \cite{103,132}. Broader datasets, such as video collections and smartphone sensing, capture diverse activities across cultures, allowing more representative benchmarks for routine analysis \cite{21,81}. Intrinsic connections are explored between individual routines and societal trends, revealing how personal habits aggregate to influence community well-being and public policy \cite{201,183}. In educational and campus contexts, studies link student routines in learning, consumption, and social life with social influence to predict broader outcomes \cite{92,100,163}. At the group level, mobility-based sensing reveals shared routines and collective formations, shedding light on how people coordinate activities, interact socially, and structure daily movements \cite{10}. 
\section{Discussion}

Our review in Section \ref{sec:Section 3} synthesizes the sensing patterns, behavioral contexts, cognitive factors, and adaptation mechanisms that characterize routine computing. Section \ref{sec:Section 4} further shows how these systems aim to support health, productivity, and well-being across diverse populations. Building on this foundation, our discussion moves beyond general technical themes to identify explicitly HCI-oriented challenges and opportunities, outlining both the major challenges and the corresponding design directions. In the following discussion, we connect the field's limitations to concrete human-oriented issues—such as interpretability gaps, feedback burden, privacy authorization, and domain-sensitive data practices—and provide design recommendations grounded in user interaction patterns, interface requirements, and human-centered design principles.

\subsection{From Activities to Intent: Context and Interpretability}

Our findings reveal that routine computing research has made significant progress in detecting low-level events and activities, yet continues to face challenges in bridging the gap from observed behaviors to contextualized interpretations of human intent. Across various sensing modalities, systems often capture \textit{what} action occurred (e.g., walking, cooking, flushing) but fail to determine \textit{why} it occurred or what meaning it holds \cite{4,6,27,28,29}. This semantic gap leads to interpretability issues: routine deviations can be flagged without clarity on whether they signal benign variation, social influences, or health decline \cite{9,12,31,73}. Similarly, ambiguities arise when identical signals map to multiple possible intentions, such as a person in bed that is asleep, resting, or unwell \cite{6,27,51,119}. Importantly, these challenges extend beyond technical limitations. The lack of contextual grounding prevents systems from supporting users' metacognition, reflection, and self-monitoring of their routines (Section \ref{sec:Section 3.3.5})—processes that are essential for helping people understand not just what they did, but why they behave as they do. Collectively, these limitations constrain the capacity of routine computing systems to move beyond pattern recognition toward meaningful explanations of lifestyle.

To advance toward interpretability, future work should prioritize the integration of richer contextual cues and hierarchical modeling of goals. Technically, combining multi-modal data streams can provide the grounding necessary to differentiate between superficially similar activities \cite{17,46,82,164}. In particular, 
egocentric sensing \cite{12,22,189,198}-likely to become more common with the adoption of AR glasses-offers a fine-grained first-person context that may help systems disambiguate user intent in ways that stationary or device-centric sensors cannot. Methods that incorporate causal and relational reasoning, such as ontology-based frameworks \cite{27}, pattern mining \cite{133}, and hierarchical temporal models \cite{158}, hold promise for connecting low-level observations with higher-order intentions and values. Moreover, longitudinal analysis of routines over weeks or months could reveal the progression and underlying causes of behavioral changes, rather than isolated deviations \cite{12,94,159}. From an interaction perspective, the key is to design transparent and interpretable mechanisms that make these system inferences cognitively accessible to users. Interfaces should reveal how individual actions are combined with contextual cues to infer intentions, allowing users to see and reflect on the causal links between their daily behaviors and system-generated interpretations. For example, if the system flags a drop in morning physical activity, the interface could display contextual information such as shorter-than-usual sleep or adverse weather conditions, helping users understand the inferred behavioral change. This aligns with the goal of context-aware assistance (Section \ref{sec:Section 4.3.2}), emphasizing that systems should not only infer intent but also surface explanations in forms that help users understand behavioral goals and potential adjustments, therefore effectively supporting human-centered applications in health, well-being, and daily life.

\subsection{Stable yet Adaptive: Balancing Personalization and Generalization in Longitudinal Modeling}

Although routine computing systems increasingly capture long-term patterns, they struggle to balance the stability of generalized models with the flexibility required for personalized adaptation. On one hand, models trained on population-level data often fail to account for individual variability in physiology, preferences, and lifestyles \cite{7,26,48,86}, limiting their utility in real-world deployment. On the other hand, highly individualized models, such as those based on unique mobility trajectories or home-specific sensor distributions, face challenges in scaling and transferring across users and domains \cite{9,14,18,47,50}. The dynamic nature of human life further complicates this tension. What constitutes a “normal" routine can shift gradually with age, health changes, or environmental adjustments \cite{9,24,27,112}. Moreover, models are often trained statically on historical data \cite{25,63,98}, without mechanisms for incremental learning to accommodate evolving habits \cite{169}. Critically, from a user-oriented standpoint, current feedback-driven personalization \cite{66,70,124,141} introduces additional challenges. Users might be required to decide when and how to report routine changes or correct misclassifications, which can impose cognitive load and may interrupt ongoing activities. These issues echo Section \ref{sec:Section 3.3} and Section \ref{sec:Section 3.4}, especially the difficulty users face in describing their own routine patterns (Section \ref{sec:Section 3.3.5}) and in expressing small deviations in ways that translate into actionable system interventions (Section \ref{sec:Section 3.4.2}). Future directions thus lie in developing longitudinal models that are both stable enough to identify enduring regularities and adaptive enough to capture personal evolution, while remaining low-burden and seamlessly integrated into users' everyday lives.

One promising approach is cross-domain fusion, where integrating temporal, spatial, physiological, and social cues can provide resilience to noise while grounding personalization in richer contexts \cite{70,76,81,83}. Transfer learning and domain adaptation methods can also allow models trained on large populations to be efficiently recalibrated for individuals \cite{48,63,69,99}. Meanwhile, interaction design plays a central role in managing feedback-driven personalization \cite{66,70,124,141}, which should reduce cognitive burden by offering lightweight feedback channels. Users often struggle to articulate why a routine has changed or what correction is needed, therefore, systems should guide them with example-based prompts, comparative visualizations (“your routine this week vs last week"), and suggestions that can be accepted or rejected with low effort. These mechanisms can transform personalization from a high-effort reporting task into a lightweight negotiation process embedded naturally into everyday interaction.

\subsection{Principles into Practice: Design for Privacy and Ethics}

Our findings show that privacy and ethical concerns remain a central barrier to the deployment of routine computing systems. Many studies highlight the tension between continuous monitoring for utility and the intrusiveness of the modalities employed, which is further intensified by the fine-grained temporal granularity required for routine computing (Section \ref{sec:Section 3.1}). Video and audio-based sensing in homes or workplaces raises concerns about intimate surveillance \cite{26,57,68,76}, while wearable and physiological monitoring can expose sensitive biometric and behavioral data when transmitted to external servers or third-party cloud platforms \cite{33,96,100}. In addition, when ambient sensors or lifelogging devices are adopted, users often express discomfort with opaque monitoring practices and unclear data handling \cite{25,84,106,154}. Privacy sensitivities are further heightened when sensing is deployed among vulnerable populations in aging and disability care (Section \ref{sec:Section 4.1}). These concerns are not only technical but also social. High opt-out rates and refusals to participate in studies manifest how privacy perceptions restrict adoption \cite{148,154}. Together, these challenges reveal a structural misalignment between the data-intensive requirements of activity recognition and the ethical imperatives of preserving consent and control in everyday environments.

A body of existing research points to multiple strategies for embedding privacy-preserving principles into practice. One approach has been to reduce invasiveness by adopting lens-free and wearable designs that limit the exposure of identifiable data \cite{144,160,166}. Other work emphasizes selective or cryptographic data transformations, such as hashed identifiers for location traces \cite{183}, anonymization protocols for group mobility \cite{145}, or user-controlled masking of sensitive streams \cite{135,151}. Beyond these system-level safeguards, another key direction is to design interfaces that allow people to understand and control the flow of temporal and contextual data without overwhelming them. For example, progressive disclosure interfaces could show what is being sensed at different levels of granularity. At the same time, preview mechanisms can simulate how routine inferences would change if certain streams were disabled. Such interaction tools allow users to actively shape the privacy boundaries of routine-aware systems rather than passively accept them, thereby enabling the sustainable integration of routine computing into daily life.

\subsection{Bridging the Data Chasm: Curation and Enrichment}

Despite the abundance of work on routine computing, the field is constrained by a persistent “data chasm." Many studies rely on small and homogeneous datasets that limit ecological validity \cite{17,20,23,49}. Even widely adopted resources such as CASAS \cite{cook2012casas} provide only partial coverage of real-world variability. Longitudinal data collection remains resource-intensive and difficult to scale \cite{13,136}, while ground-truth labeling is often inconsistent or relies on manual inspection\cite{19,26,69}. In-the-wild data introduces further complexity, including noisy, sparse, and unbalanced signals \cite{7,116,165}, sensor placement inconsistencies \cite{50,111}, and ethical constraints that require data collection from healthy subjects rather than target groups \cite{29,92}. These constraints not only hinder generalization across devices and environments \cite{63,132}, but also obscure the detection of rare or anomalous events that are often artificially induced rather than naturally observed \cite{16,78}. Therefore, these challenges underscore the gap between experimental feasibility and the demands of robust and scalable routine computing, particularly in supporting goals such as translating individual routines into population-informed insights (Section \ref{sec:Section 4.4}).

In response, recent work points toward a layered set of strategies for bridging this chasm. One orientation is toward multi-dataset training and cross-source integration, which can mitigate modality-specific biases and support broader generalization\cite{63,183}. While real-world data remains scarce, synthetic dataset generation and co-occurrence analysis are utilized. Innovations in labeling protocols, such as standardized schemes\cite{1} or sophisticated self-reporting applications\cite{69}, suggest pathways to improve consistency in ground truth acquisition. Apart from algorithmic advances\cite{7,116}, future work should also emphasize human-group-sensitive curation principles, such as collecting data from clinically or socially relevant populations \cite{29,92}. There is also an opportunity for HCI to support more sustainable practices by designing tools that help participants contribute longitudinal data with less burden. Examples include self-report aids, micro-interaction logging tools, or interfaces that summarize routine changes and allow users to validate or annotate them retroactively. Moreover, egocentric life-logging systems can be conceptualized to gain higher-quality personal routine data. Such platforms can be designed by fusing various sensing modalities with privacy-aware mechanisms to protect sensitive information and ensure transparency. All these directions suggest a shift from isolated datasets and ad hoc evaluation toward an integrated and enriched resource ecosystem. Addressing this data chasm is not merely a technical prerequisite but a foundational step for enabling routine computing to move from controlled prototypes to sustainable, real-world systems that genuinely serve human users.

\section{Limitations}
A limitation of this review is that our database scope was restricted to ACM and IEEE venues, which historically anchor routine computing through HCI, ubiquitous computing, sensing, and AI research. While preliminary scoping of Springer, ScienceDirect, and NIH yielded primarily medical or clinically oriented monitoring studies with limited computational routine modeling relevance, future work may broaden coverage across these sources to capture cross-disciplinary perspectives and further strengthen comprehensiveness.

\section{Conclusion}

This review synthesized 203 studies to chart the evolving landscape of routine computing for daily life. We identified four key dimensions that characterize how routines are sensed and understood—temporal granularity, interactional contexts, cognitive-psychological factors, and the dynamics of variability and disruption. We further organized the goals of routine computing into domains spanning support for aging and disability care, the promotion of health practices and habits, adaptive assistance through reflection and prediction, and the extension from individual routines to population-level insights. Informed by these results, our discussion is structured around two overarching themes: persistent challenges and future prospects, including enhancing interpretability of routine models, balancing personalization with generalizability in longitudinal settings, embedding privacy and ethical principles into system design, and advancing data curation and enrichment practices. Together, these directions outline a research agenda that positions routine computing as a critical frontier for human-centered HCI innovation in everyday life.

\begin{acks}

This paper is supported by the National Key R\&D Program under Grant No. 2024YFB4505500 \& 2024YFB4505503, National Natural Science Foundation of China under Grant No. 62472244, Qinghai University Research Ability Enhancement Project under Grant No. 2025KTSA05, Ant Group Research Fund, Tsinghua University Initiative Scientific Research Program and Undergraduate Education Innovation Grants of Tsinghua University.
\end{acks}
\bibliographystyle{ACM-Reference-Format}
\bibliography{ref}


\end{document}